\documentclass[12pt]{article}
\usepackage{amsmath,amsfonts,amssymb,latexsym}

\numberwithin{equation}{section}
\def\be{\begin{equation}}
\def\ee{\end{equation}}
\def\bq{\begin{eqnarray}}
\def\eq{\end{eqnarray}}
\def\beq{\begin{eqnarray*}}
\def\eeq{\end{eqnarray*}}

\def\g{\gamma}
\def\G{\Gamma}

\def\l{\lambda}

\begin{document}
\title{\huge{Asymptotic Poincar\'e compactification and finite-time singularities}}
\author{\Large{Spiros Cotsakis\footnote{\texttt{email:\thinspace skot@aegean.gr}} } \\
{\normalsize Research Group of Geometry, Dynamical Systems and
Cosmology}
\\ {\normalsize University of the Aegean}\\ {\normalsize Karlovassi 83 200, Samos,
Greece}}
\maketitle
\begin{abstract}
\noindent We provide an extension of the method of asymptotic decompositions of vector fields with finite-time singularities by applying the central extension technique of Poincar\'e to the dominant part of the vector field on approach to the singularity. This leads to a bundle of fan-out asymptotic systems whose equilibria at infinity govern the dynamics of the asymptotic solutions of the original system. We show how this method can be useful to describe a single-fluid isotropic universe at the time of maximum expansion, and discuss possible relations of our results to structural stability and non-compact phase spaces.
\end{abstract}

\section{Introduction}	
Sufficient conditions for the global existence of general relativistic spacetimes imply their physical duration for an infinite proper time, as shown in Refs. \cite{ycb-cot02,cot04}. The implied singular universes (in the contrapositive direction) will necessarily have \emph{finite-time} singularities, and their classification problem is a highly non-trivial matter already in the isotropic category, cf. \cite{ck05,ck07,cot07}. These include for instance, apart from the standard singular universes of relativistic cosmology, sudden singularities \cite{ba04}, dark energy ones \cite{noj1}, singular universes with interacting fluids \cite{sc12a}, universes with higher order gravity corrections \cite{K1}, and brane models \cite{kl1,kl2,nima}.

A popular method to represent infinity in general relativity is Penrose's conformal method (cf. Ref. \cite{pen}) wherein the overall structure of a physical spacetime is conformally changed so that its infinitely remote regions become the boundary of a new, unphysical spacetime. Infinity is then classified according to the behaviour of the various geodesics of the new spacetime near its boundary. Until recently, the conformal method was the only useful method to study the 'structure of infinity' in relativistic field theories. In finite dimensions, the corresponding idea to analyze the global structure of solutions especially for planar systems stereographically is due to Bendixson, cf. Ref. \cite{b}.

The method of asymptotic splittings (cf.  \cite{split} and refs. therein) offers another  asymptotic representation of solutions to the dynamical equations near infinity (that is where some component of the solution diverges), and it is possible to describe it in  three main steps as follows. First, we recall some simple definitions. We shall work on open subsets or
$\mathbb{R}^{n}$, although our results can be extended without difficulty to any differentiable manifold $\mathcal{M}^{n}$.
 We shall use interchangeably the terms vector field $
f:\mathcal{M}^{n}\rightarrow \mathcal{TM}^{n}$ and dynamical system defined
by $f$ on $\mathcal{M}^{n}$, $\dot{x}=f({x})$, with $(\cdot )\equiv d/dt$. Also, we will use the terms `integral curve'
${x}(t,{x}_{0})$ of the vector field ${f}$ with initial
condition ${x}(0)={x}_{0}$, and 'solution' of the associated
dynamical system ${x}(t;{x}_{0})$ passing through the point $%
{x}_{0}$, with identical meanings. We say that the
system $\dot{{x}}={f}({x})$ (equivalently, the vector
field ${f}$) has a \emph{finite-time singularity} if there exists a $%
t_{s}\in \mathbb{R}$ and a ${x}_{0}\in \mathcal{M}^{n}$ such that
for all $M\in \mathbb{R}$ there exists an $\delta >0$ such that
$||{x}(t;{x}_{0})||_{L^{p}}>M, $
for $|t-t_{s}|<\delta $. Here ${x}:(0,b)\rightarrow \mathcal{M}%
^{n}$, ${x}_{0}={x}(t_{0})$ for some $t_{0}\in (0,b)$, and $%
||\cdot ||_{L^{p}}$ is any ${L^{p}}$ norm. We say that the vector field has
a \emph{future} (resp. \emph{past}) singularity if $t_{s}>t_{0}$ (resp. $%
t_{s}<t_{0}$). Note also, that $t_{0}$ is an arbitrary point in the
domain $(0,b)$ and may be taken to mean `now'. Alternatively, we may set $%
\tau =t-t_{s}$, $\tau \in (-t_{s},b-t_{\ast })$, and consider the
solution in terms of the new time variable $\tau $, ${x}(\tau ;%
{x}_{0})$, with a finite-time singularity at $\tau =0$.
We see that
for a vector field to have a finite-time singularity there must be at least
one integral curve passing through the point ${x}_{0}$ of $\mathcal{M}^{n}$ such that at least one of its ${L^{p}}$ norms diverges at $t=t_{s}$,  and we write
\begin{equation}
\lim_{t\rightarrow t_{s}}||{x}(t;{x}_{0})||_{L^{p}}=\infty,
\label{sing2}
\end{equation}
to denote the behaviour of the solution on approach to the  finite-time singularity at $t_{s}$.

One of the most interesting problems in the theory of singularities of
vector fields is to find the structure of the set of points ${x}_{0}$
in $\mathcal{M}^n$ such that, when evolved through the dynamical system
defined by the vector field, the integral curve of ${f}$ passing
through a point in that set satisfies property (\ref{sing2}). The evolutionary, geometric nature of this set especially near singularities of the field, is of prime interest in this paper. In the method of asymptotic splittings,
we imagine that on approach to a finite time singularity, ${f}$ decomposes into a \emph{dominant} part ${f}^{\,(0)}$ and another, \emph{subdominant} part (consisting possibly of more than one additive components ${f}^{\,(\mu)}, \mu=1,\cdots,k$), as follows:
\begin{equation}
{f}={f}^{\,(0)}+{f}^{\,(1)}+\cdots +{f}^{\,(k)}.
\label{dec1}
\end{equation}
This asymptotic decomposition is highly non-unique. Then, in the first step, for each particular decomposition we drop all terms after ${f}^{\,(0)}$ and replace the exact equation  $\dot{{x}}={f}({x})$  by the \emph{asymptotic equation}
\be
\dot{{x}}={f}^{\,(0)}({x}).
\ee
Repeating this for all possible decompositions of ${f}$, we end up with a bundle of asymptotic dynamical systems to work with.
In Step two, we look for scale-invariant solutions (called dominant balances) in each asymptotic system. Each such system may contain  many possible dominant balances resulting in various possible dominant behaviours near the singularity. Balancing each asymptotic system, requires a careful asymptotic analysis of the subdominant parts in the  various  vector field decompositions.
In the last step, we check the overall consistency of the approximation scheme and build asymptotic solutions for each acceptable balance in a term-by-term iteration procedure that ends up with a formal series expansion representation of the solutions.
When the whole procedure is completed, we have finally constructed formal series developments of particular or general solutions of the original system of equations, valid in a local neighborhood of the finite-time singularity. From such expansions we can  deduce all possible dominant modes of approach of the field to the singularity,
  decide on the generality of the constructed solutions and also
   determine the size and part of the space of initial data that led to such a solution.
However, this method  has two shortcomings:
\begin{enumerate}
  \item It does not give any information about the \emph{qualitative} behaviour of individual orbits
  \item It does not distinguish between the behaviour of those solution components  tending to $+\infty$ from those ones  tending to $-\infty$.
\end{enumerate}

The qualitative method of central projections and compactifications due to Poincar\'e is usually presented, cf. Refs. \cite{lef,per,dum,mei}, as suitable only for homogeneous vector fields and singularities at infinity, that is when we have a homogeneous vector  field diverging as $t\rightarrow\infty.$ The purpose of this paper is to emphasize the fact that this method, when considered from the viewpoint of asymptotic splittings, is really suitable for any \emph{weight-homogeneous} vector field that blows up at a \emph{finite} time (or at infinity). In this way, we may suitably adapt the ingenious method of Poincar\'e's to study the finite-time  singularities that arise  in cosmology and relativistic field theory, wherein a healthy dose of problems having precisely this kind of vector field and asymptotic structure appear with phase spaces of any finite  number of dimensions. It is hoped that this will provide a useful alternative to the stereographic techniques of Bendixson and Penrose mentioned above.

In this paper, we revisit the method of central projections and use it as a suitable complement and development of the method of asymptotic splittings, in order to decide about the \emph{qualitative} behaviours of the asymptotic orbits that result from the various possible asymptotic decompositions. In the next section, we give the details of the method of central projection for one-dimensional systems, and describe the state of maximum expansion for a universe with a single fluid in general relativity, a problem known not to admit a complete description with standard methods. In Section 3, we describe the amalgamation of asymptotic splittings and central projections in two dimensions, and in the last section we discuss various issues that lie ahead and result from the approach taken in this paper.

\section{The one-dimensional case}
Let us consider the one-dimensional system $\dot{x}=f(x)$ and assume that there is a finite time $t_s$ such that  the solution $x$ has a singularity, $x(t_s)=+\infty$, or $x(t_s)=-\infty$. In general relativity (which is the principal area of application we have in mind for the techniques discussed here), one is accustomed to view such singularities as following from other plausible physical and geometric hypotheses - the so-called singularity theorems, cf. \cite{he73,ycb,sc12e}. We imagine the one-dimensional phase space $\G=\{x:x\in\mathbb{R}\}$ of the system as sitting at the $(X,1)$ (cotangent) line of an $(X,Z)$ plane, and we let $x$ be any particular state of the system on this line (so that $x=X$ in this case). The Poincar\'e central projection $\textrm{pr}_c$ is a map from $\G$ to the upper semi-circle $\mathcal{S}_+^1$ sending each phase point to a point on the circle such that
\be
\textrm{pr}_c:\G\rightarrow\mathcal{S}_+^1:x\mapsto \theta ,\quad\textrm{with}\quad x=\cot\theta.
\ee
In this way, the two possible infinities of $x$ at the two ends of $\G$ are bijectively separated (note that this cannot happen with a stereographic map like Penrose's),
\bq
x=+\infty&\rightarrow&\theta=0\\
x=-\infty&\rightarrow&\theta=\pi,
\eq
the new phase space, the so-called \emph{Poincar\'e's circle}, $\mathcal{S}_+^1$ is now compact, $\theta\in [0,\pi]$, and the original system $\dot{x}=f(x)$  reads,
\be
\dot\theta=-\sin^2\theta f(\cot\theta )\equiv g(\theta).
\ee
So past singularities are met as $\theta\downarrow 0$, while future ones are at $\theta\uparrow\pi$. Let us suppose for the sake of illustration that we are interested in  the asymptotic  properties of the solution near a \emph{past} singularity. In the next step of the Poincar\'e central projection method, we are after a singular \emph{asymptotic} system on the compactified phase space $\mathcal{S}_+^1$. According to the method of asymptotic splittings,  the field $f$ is asymptotically decomposed into two parts, a dominant one $f^{(0)}$, and another subdominant, $f^{(\textrm{sub})}$, i.e., $f=f^{(0)}+f^{(\textrm{sub})}$, in a highly non-unique manner. We take any particular decomposition for which
$
f^{(0)}=ax^M, M\in\mathbb{Q},
$
and asymptotically as $\theta\downarrow 0$ we find that
$
g(\theta)=-a\theta^{2-M},
$
that is we arrive at the \emph{singular asymptotic system}
\be
\dot\theta=-a\theta^{2-M}.
\ee
As in the method of asymptotic splittings, we need to repeat the whole procedure below for all possible decompositions. Notice that this system is only valid locally around the singularity at $\theta =0$, not everywhere on the compactified phase space $\mathcal{S}_+^1$. The last step is now to obtain a \emph{singularity-free} system valid locally around the singularity. For this purpose, one changes the time $t$ to a new one, $\tau$, given by $d/dt=\theta^{1-M}d/d\tau$, to arrive at the \emph{complete asymptotic system} defined on the Poincar\'e circle,
\be
\dot\theta=-a\theta,
\ee
and valid near $\theta=0.$  We conclude that the system has a past attractor when $a>0$, otherwise all orbits are repelled near the singularity.

To assess the merits of this method, we revisit the evolution  of the single-fluid FL models in general relativity with a linear equation of state $p=(\g-1)\rho$, studied  in Ref. \cite{we97}, pp. 58-60. Introducing the \emph{density parameter} $\Omega$ defined in terms of the fluid density $\rho$ and Hubble rate $H$ by $\Omega=\rho/3H^2$, the evolution of the system is governed by the equation
\be
\dot\Omega=-(3\g-2)(1-\Omega)\Omega,
\ee
and it is well known that for closed models, after a finite time interval, at the instant of maximum expansion, $\Omega$ blows up to $+\infty$, thus making the description of the whole evolution incomplete (cf. Ref. \cite{we97}, p. 60). To describe this, we choose the decomposition having $f^{(0)}=a\Omega^2$, where $a=3\g-2$ and $M=2$. The resulting  asymptotic system is $\dot\theta =-(3\g-2)$, and taking $\theta=0$ at $t=t_{\textrm{max}}$, we find that $\theta =-(3\g-2)(t-t_{\textrm{max}})$. The central projection gives $\Omega=\cot\theta=\cot(-(3\g-2)(t-t_{\textrm{max}}))$, near the point of maximum expansion.

\section{Dimension 2}
For homogeneous vector fields, the various reductions involved in central projections are described in Refs. \cite{lef,per,dum,mei} and will not be repeated here. Below, as in the method of asymptotic splittings, we focus on weight-homogeneous vector fields and finite-time singularities exclusively. Any $\mathcal{C}^1$ function $h$ is called \emph{weight-homogeneous} if we can find a $d\in\mathbb{R}$ and a nonzero vector $w\in\mathbb{R}^n$ such that for all numbers $t$, we have
\be
h(t^{w_1}x_1,t^{w_2}x_2,\cdots,t^{w_n}x_n)=t^dh(x_1,x_2,\cdots,x_n).
\ee
We then speak of the \emph{weight vector} $w=(w_i)$ and the \emph{weighted degree} $\textrm{deg}(h,w)=d$ of the function $h$. For instance, the function $f_1=2xy$ is weight-homogeneous with weight vector $w=(-1,-1)$ and weighted degree $d=-2$, and $f_2=4y^2-x^2$ is also weight-homogeneous with the same $w$ and $d$. When $w=(1,\cdots,1)$, we say that $h$ is a \emph{homogeneous} function.

A \emph{weight-homogeneous vector field} $f=(f_1,\cdots,f_n)$ is one for which we can find a a \emph{common} weight vector $w$ for the $f_i$'s. When the components $f_i, i=1,\cdots,n,$ have weighted degrees $d_i=\textrm{deg}(f_i,w), i=1,\cdots,n,$, we say that the vector field $f$ has \emph{weighted degree} $\textrm{deg}(f,w)=d$ with $d=(d_i)$. For example, the vector field $f=(4y-x^2/2,xy+x^3)$ has weight vector $w=(-1,-2)$ and weighted degree $d=(-2,-3)$.

An important subclass of the weight-homogeneous vector fields are the \emph{scale-invariant} vector fields, for which  their weight vectors and degrees are related by,
\be
d=w-1,
\ee
with $1$ here meaning the vector having $1$ in every slot (hence, $d_i=w_i-1$). Thus the degree $d$ of a scale-invariant vector field measures  its failure from being exactly homogeneous. An obvious property of any scale-invariant field $f$ with weight vector $p$, is that  the `scale-invariant' system $\dot{x}=f$ admits a solution of the form $x=at^p, a\in\mathbb{C}^n$ (in this notation, $x_i=a_i t^{p_i}$ in components), when the algebraic system  $a\cdot p=f(a)$ has a nontrivial solution (that is some component of $a$ is nonzero). In this case, the solution  $x=at^p$ is invariant under the scaling $t\rightarrow \l t, x\rightarrow \l^p x$.

In the method of asymptotic splittings, we look for \emph{weight-homogeneous decompositions}, that is vector field decompositions  of the form (\ref{dec1}) such that the following conditions hold:
 \begin{enumerate}
 \item all vector field `components' ${f}^{\,(j)}$ in the sum are weight-homogeneous with the same weight vector $w$,
 \item their weighted degrees are given by $d^{(j)}=w-j-1$, and $d^{(0)}>d^{(1)}>\cdots>d^{(k)}$,
 \item the most nonlinear, additive component ${f}^{\,(0)}$ is scale-invariant.
 \end{enumerate}
For the sake of illustration, we shall confine our attention to planar systems of the form
\be\label{basic}
\dot x=P(x,y),\quad \dot y=Q(x,y),
\ee
and we shall assume that this system has a finite-time singularity at time $t_s$, as in the definition (\ref{sing2}) (everything we do generalizes easily to any finite dimension).  As in the 1-dimensional case of the previous section, since the original system (\ref{basic}) is incomplete, we are eventually after a complete, `asymptotic' system defined on the 2-dimensional sphere $\mathcal{S}^2$. Taking the phanar phase space $(x,y)$ to be the $Z=1$ plane of a new 3-space with coordinates $X,Y,Z$, the embedding
\be\label{prc}
\mathrm{pr}_c:\mathbb{R}^2\rightarrow\mathcal{S}_+^2:(x,y)\mapsto (X,Y,Z): x=X/Z,\quad y=Y/Z,\quad X^2+Y^2+Z^2=1,
\ee
interprets any blow up in the solution vector $(x,y)$, $x,y\rightarrow\infty$, as the limit  $Z\rightarrow 0$. (We note that any sign in the $x,y$'s is taken care of by the signs of the new $X,Y$ variables, and so we can limit ourselves on the positive semi-sphere, that is consider only the direction $Z\downarrow 0$.) Thus infinity in $x,y$ is now (at) the $Z=0$ \emph{circle of infinity}, $X^2+Y^2=1$. This is
the first step of the so-called Poincar\'e compactification for the system (\ref{basic}) wherein the embedding  $\textrm{pr}_c$ is  the  \emph{central projection transform}, and sends a planar phase point $(x,y)$ to a point of the semi-sphere (the 'Poincar\'e sphere') $\mathcal{S}_+^2$. Under the map (\ref{prc}), the system (\ref{basic}) becomes the \emph{singular} system \cite{lef,per,dum,mei},
\bq
\dot X&=&Z\left((1-X^2)P-XYQ\right)\nonumber\\
\dot Y&=&Z\left(-XYP+(1-Y^2)Q\right)\label{basic2}\\
\dot Z&=&-Z^2(XP+YQ)\nonumber.
\eq
We next aim to  obtain asymptotically, through a series of transforms, first a reduction of the system (\ref{basic2}) that will be complete and valid on $\mathcal{S}_+^2$, and secondly another reduction valid precisely \emph{only along} the circle of infinity. We suppose we have used the method of asymptotic splittings and ended up with the  spectrum of all possible asymptotic decompositions of the form (\ref{dec1}) of the  vector field $(P(x,y),Q(x,y))$. We shall work with a particular weight-homogeneous  decomposition that has dominant part given by
\be
{f}^{\,(0)}=(P^{\,(0)},Q^{\,(0)}),
\ee
and we suppose that $a=(a_i), i=1,2,$ is the weighted degree of $f^{(0)}$. We then define,
\be\label{m}
M=\max \{|a_i|,i=1,2\},
\ee
and using this, we can introduce the \emph{time transform} $t\rightarrow\tau (t)$ (monotone when $Z>0$),
\be
\frac{d}{dt}=Z^{1-M}\frac{d}{d\tau},
\ee
so that the singular system (\ref{basic2}) becomes the following system for the dominant part of the vector field $(P,Q)$ ($(\,')\equiv d/d\tau$),

\textsc{Intermediate system:}
\bq
X'&=&Z^M\left((1-X^2)P^{\,(0)}-XYQ^{\,(0)}\right)\nonumber\\
Y'&=&Z^M\left(-XYP^{\,(0)}+(1-Y^2)Q^{\,(0)}\right)\\
Z'&=&-Z^{M+1}\left(XP^{\,(0)}+YQ^{\,(0)}\right)\nonumber.
\eq
There is an important difference with respect to the homogeneous case treated in Refs. \cite{lef,per,dum,mei}, in that we consider only the dominant part of the vector field asymptotically, and this will be different for each particular asymptotic decomposition of the field. The `subdominant' part is truly subdominant asymptotically because of the validity of the \emph{subdominance condition}, cf. \cite{split},
\be
\lim_{t\rightarrow 0}\frac{f^{\textrm{sub}}(at^{p})}{t^{p-1}}=0.
\ee
In a sense, the standard presentation of the  Poincar\'e method corresponds to the all-terms-dominant decomposition presently (the only possible one when the field is homogeneous). We can further make this system a complete asymptotic system valid on  $\mathcal{S}_+^2$ by introducing the \emph{sphere vector field} $(P_{\mathcal{S}_+^2},Q_{\mathcal{S}_+^2})(X,Y,Z)$ defined by the forms,
\bq
P_{\mathcal{S}_+^2}(X,Y,Z)&=&Z^M P^{\,(0)}\left(\frac{X}{Z},\frac{Y}{Z}\right)\\
Q_{\mathcal{S}_+^2}(X,Y,Z)&=&Z^M Q^{\,(0)}\left(\frac{X}{Z},\frac{Y}{Z}\right).
\eq
Then the intermediate system above becomes one valid on $\mathcal{S}_+^2$ without singularities, namely,

\textsc{Complete asymptotic system on $\mathcal{S}_+^2$:}
\bq\label{sphere system}
X'&=&(1-X^2)P_{\mathcal{S}_+^2}-XYQ_{\mathcal{S}_+^2}\nonumber\\
Y'&=&-XYP_{\mathcal{S}_+^2}+(1-Y^2)Q_{\mathcal{S}_+^2}\\
Z'&=&-Z\left(XP_{\mathcal{S}_+^2}+YQ_{\mathcal{S}_+^2}\right)\nonumber.
\eq
We note that in this system, any additive  terms proportional to $Z$ are asymptotically negligible on approach to the singularity, and so this system asymptotically becomes one valid on the $(X,Y)$-plane \emph{along} the circle of infinity, $\mathcal{S}^1:X^2+Y^2=1$, that is we have

\textsc{Asymptotic system on circle of infinity:}
\bq
X'&=&-Y(XQ_{\mathcal{S}^1}-YP_{\mathcal{S}^1})\label{sys1}\\
Y'&=&X(XQ_{\mathcal{S}^1}-YP_{\mathcal{S}^1})\label{sys2},\quad\textrm{on the circle}\quad X^2+Y^2=1.
\eq
Here we have introduced the \emph{circle vector field} $(P_{\mathcal{S}^1},Q_{\mathcal{S}^1})(X,Y)$ given by
\bq
P_{\mathcal{S}^1} (X,Y)&=&P_{\mathcal{S}_+^2}(X,Y,0)\label{c1}\\
Q_{\mathcal{S}^1} (X,Y)&=&Q_{\mathcal{S}_+^2}(X,Y,0)\label{c2}.
\eq
Some remarks about the system (\ref{sys1}-\ref{sys2}) defined on the Poincar\'e circle of infinity are in order. First, through the series of transforms defined above, for any weight-homogeneous vector field $f$ with a finite-time singularity we have found that infinity in the original variables $x,y$ is  now  an invariant manifold (circle), with the dynamics near infinity determined by the constrained 2-dimensional system (\ref{sys1}-\ref{sys2}), effectively 1-dimensional, on the equator. Secondly, for any given system of the form (\ref{basic}), there correspond precisely the same number of systems of the form (\ref{sys1}-\ref{sys2}) as the number of \emph{admissible} asymptotic decompositions of the original vector field (\ref{basic}). We note that an asymptotic splitting is admissible provided we come up eventually with a valid asymptotic solution in the form of a formal series. In particular, we have to find the Kowalevskaya exponents of the various dominant balances of the given decomposition and check that each one of these leads to a consistent asymptotic scheme as in \cite{split}.

The equilibria of (\ref{sys1}-\ref{sys2}) correspond to the `equilibria at infinity' of the flow of (\ref{sphere system}) on  the Poincar\'e sphere $\mathcal{S}_+^2$, and they are precisely the points where
\bq
XQ_{\mathcal{S}^1}-YP_{\mathcal{S}^1}&=&0\label{eq1}\\
X^2+Y^2&=&1,\label{eq2}
\eq
with $(P_{\mathcal{S}^1},Q_{\mathcal{S}^1})(X,Y)$ being the circle vector field introduced above. Therefore  the study of the flow of the original system (\ref{basic}) near its finite-time singularities is equivalent to that of the behaviour of the orbits of the asymptotic system (\ref{sys1}-\ref{sys2}) near its equilibria given by (\ref{eq1}-\ref{eq2}). Because of the circle condition (\ref{eq2}), these equilibria are of the form $(X,Y,0)$ and come in as antipodal pairs distributed along the circle of infinity, hence it is only necessary to study the flow along the equilibria having $X>0$, and secondly those having $Y>0$ (in particular, they cannot both be zero). The flow will be topologically equivalent (with direction reversed) at the antipodal points if  $M$, defined in (\ref{m}), is odd (even) as in the homogeneous case, cf. \cite{per,mei}.

To study the flow near such equilibria, we follow  a `fan-out' technique, described in Refs. \cite{per,mei} for the all terms dominant case. Suppose we have an equilibrium, solution of (\ref{eq1}-\ref{eq2}), with $Y>0$. We define the \emph{fan-out map}
\be \xi=\frac{X}{Y},\quad\zeta=\frac{Z}{Y}\ee
which spreads any small neighborhood of that equilibrium on the circle back out onto the tangent plane to the Poincar\'e sphere at $Y=+1$. Then the geometric dynamics defined by the asymptotic system (\ref{sys1}-\ref{sys2}) near this equilibrium is transformed into one described by the fan-out system near its corresponding equilibrium in terms of the new variables $\xi,\zeta$:

\textsc{Fan-out system for $Y>0$:}
\bq
\dot\xi&=&\zeta^M P_{\mathcal{S}^1}\left(\frac{\xi}{\zeta},\frac{1}{\zeta}\right)-\xi\zeta^M Q_{\mathcal{S}^1}\left(\frac{\xi}{\zeta},\frac{1}{\zeta}\right)\\
\dot\zeta&=&=\zeta^{M+1}Q_{\mathcal{S}^1}\left(\frac{\xi}{\zeta},\frac{1}{\zeta}\right).
\eq
This is valid in a neighborhood of the point $(0,+1,0)$. Similarly, there is a completely analogous fan-out reduction for any equilibrium having $X>0$. The final result is:

\textsc{Fan-out system for $X>0$:}
\bq
\dot\eta&=&\l^M Q_{\mathcal{S}^1}\left(\frac{1}{\l},\frac{\eta}{\l}\right)-\eta\l^M P_{\mathcal{S}^1}\left(\frac{1}{\l},\frac{\eta}{\l}\right)\\
\dot\l&=&=-\l^{M+1}Q_{\mathcal{S}^1}\left(\frac{1}{\l},\frac{\eta}{\l}\right),
\eq
where the fan-out map in this case is $\eta=Y/X,\l=Z/X$. Note that in both fan-out systems we use the asymptotic decompositions of the field.

We conclude that the dynamics of the original system near its singularities is determined by the nature of the equilibria of the corresponding fan-out systems given above, and these can be studied using standard dynamical systems methods. In particular, if these turn out to be hyperbolic, then the situation will result in a relatively easy problem, whereas when we have non-hyperbolic equilibria, the dynamics in phase space `at infinity' will be more involved and the resulting global phase portraits will look more complicated.

\section{Discussion}
In this paper we have described a combination of the techniques of asymptotic splittings and central extensions and provided an asymptotic from of the Poincar\'e compactification suitable for the analysis of finite-time singularities. This applies to weight-homogeneous vector fields and their decompositions near a time when (some component of) the solution blows up. This description will be useful in treating problems in mathematical cosmology and other fields where such situations naturally arise.

Although we have described the asymptotics of the Poincar\'e central extension method in dimensions 1 and 2, there is nothing to prevent these results from being valid in any finite dimension. The 3-dimensional case of a homogeneous vector field with a singularity as $t\rightarrow\infty$ is described in Ref. \cite{per}, and it is straightforward to formulate the results of the present paper in this context. In practise, however, we do not expect to deal with problems in dimensions higher than four, except in rare circumstances. For an abstract formulation in general dimensions on manifolds, we refer to \cite{go}.

The study of infinity in the context of the present paper opens the way for related structural stability studies and applications to mathematical cosmology where such issues have especially important physical and geometric interpretations, for instance a characterization of instabilities through the possible non-hyperbolicity of equilibria at infinity. Although, Peixoto's theorem  for structurally stable vector fields concerns \emph{compact} manifolds, there is a suitable extension of structural stability to the non-compact case, (for a modern introduction to these results, cf. Ref. \cite{per}.

Such results are of great importance in the present context, for although having the field blowing up only as $t\rightarrow\infty$ leads possibly to compact phase spaces and so stability may be natural for such vector fields when they are further compactified on the sphere, the presence of finite-time singularities and the related possibility for orbits to escape to infinity in a finite time governed only by the dominant part of the vector field, opens real possibilities for the influence of non-compact phase spaces. This is eventually related to the possible validity of the index theorem under the precise conditions of the problem analyzed here, and it is perhaps interesting that known results in this direction, cf. \cite{cima}, are not directly relevant to the issues considered in this paper.

%\section*{Acknowledgements}

\end{document}